\documentstyle[12pt]{article}
\begin{document}
\centerline{\bf CP Violation in $t\rightarrow W^+b$ Decays in Two-Higgs
Doublet Model}
\baselineskip=8truemm
\vskip 2.0truecm
\centerline{Tsutom HASUIKE$,^{a),}$\footnote{e-mail: hasuike@anan-nct.
ac.jp} \ Toshihiko HATTORI$,^{b),}$\footnote{e-mail: hattori@ias.
tokushima-u.ac.jp} }
\centerline{Toshio HAYASHI$,^{c),}$\footnote{e-mail: hayashi@ed.
kagawa-u.ac.jp} \ and \ Seiichi WAKAIZUMI$ ^{d),}$\footnote{e-mail: 
wakaizum@medsci.tokushima-u.ac.jp} }
\vskip 0.8truecm
\centerline{\it a) Department of Physics, Anan College of Technology, 
Anan 774, JAPAN}
\centerline{\it b) Institute of Theoretical Physics, University of Tokushima,
Tokushima 770, JAPAN}
\centerline{\it c) Department of Physics, University of Kagawa, 
Takamatsu 760, JAPAN}
\centerline{\it d) School of Medical Sciences, University of Tokushima,
Tokushima 770, JAPAN}
\vskip 2.0truecm
\centerline{\it ABSTRACT}
\vskip 0.5truecm

Due to the large mass of top quark, CP violation in the top-quark decay is 
sensitive to the interactions mediated by Higgs bosons. We consider CP 
violation in $t\rightarrow W^+b$ decay by calculating consistently in 
unitary gauge the CP-violating up-down asymmetry of the leptons from 
$W$ boson decays in $t \rightarrow W^+b$, defined by 
Grz\c{a}dkowski and Gunion, in the two-Higgs doublet model with 
CP-violating neutral sector. The asymmetry is shown to be at most 
of the order of $(1-3)\times 10^{-4}$ for $\tan\beta = 1.0$, where 
$\tan\beta$ is the ratio of vacuum expectation values for the 
two neutral Higgs bosons.
\newpage

\leftline{\bf 1.Introduction}
\vskip 0.4truecm
\noindent
The Kobayashi-Maskawa mechanism of CP violation is simple and very 
beautiful\cite{Kobayashi73}. However, the origin is not yet finally verified 
by experiments, 
though that in the neutral kaon system seems to be explained well 
by the mechanism, and it is expected to be tested by B-meson decays 
in the coming B factories.

On the other hand, CP violation in the ordinary top-quark decay and 
top-pair production is estimated to be very small in the 
Kobayashi-Maskawa Standard Model[2-3]. This fact has given us an 
opportunity of investigating the non-standard origin of CP violation 
in the top decay and in the top-pair production[4-11].

There are some advantages in studying CP violation in the top-quark 
system; first, because of its large mass ($m_t\sim 180$ GeV\cite{mtmass}), 
the top-quark decays very fast, its lifetime being shorter than $10^{-23}$s
\cite{Bigi86}, so that the top would decay before it hadronizes and we do 
not have to worry about the hadronic effects, and secondly, for the same 
reason, the top decay interactions are sensitive to Higgs boson exchanges. 
This nature of top-quark has lead to many works of CP violation in 
the multi-Higgs doublet model[7,14-16]. Interesting among the Higgs 
models is two-Higgs doublet model of type II\cite{Gunion89}, in which 
CP violation is caused explicitly or spontaneously in its neutral sector
\cite{Branco85}. Many authors have studied a variety of CP-violating 
observables in the top-quark productions\cite{Schmidt92} and 
decays[7,16] in this model.

In this paper, we investigate CP-violation in $t\rightarrow W^+b$ 
decay by calculating the up-down asymmetry of the leptons from 
$W$ decay with respect to the $t\rightarrow W^+b$ decaying plane, 
which is formed by $t$ production in $e^+e^-$ or hadron colliders, 
in the two-Higgs doublet model with CP-violating neutral sector. In the 
next section, the asymmetry of the lepton distributions is briefly 
expressed by the non-standard $t\ b W$-vertex form factor. In section 3, 
two-Higgs doublet model is described in order to express the 
CP-violating part of the form factor as one-loop effects with CP-violating 
neutral scalar exchanges, and numerical results of the CP-violating 
asymmetry  are given. Section 4 includes conclusions and discussions.
\vskip 1truecm

\leftline{\bf 2. Asymmetry of lepton distributions in top decays}
\vskip 0.4truecm
\noindent
We here investigate CP violation in $t\rightarrow W^+b$ decays by the 
lepton distribution asymmetry from $W^+$ decay defined by 
Grz\c{a}dkowski and Gunion\cite{Grzad92}, assuming that the top 
quark is produced in $t\bar t$ pairs in $e^+e^-$ collision, for example.

The phase space for $e^+e^-\rightarrow t\bar t \rightarrow W^+W^-b\bar b 
\rightarrow \ell^+\ell^-\nu\bar \nu b\bar b$ is given by\cite{Grzad93}
\begin{eqnarray}
{\rm d}\Phi &=& (2\pi)^{-4}{\rm d}s_t{\rm d}s_{\bar t}{\rm d}s_{W^+}
{\rm d}s_{W^-}{\rm d}\Phi (e^+e^-\rightarrow t\bar t ){\rm d}\Phi 
(t\rightarrow W^+b){\rm d}\Phi (W^+\rightarrow \ell ^+\nu )  \qquad 
\nonumber   \\
& &\times{\rm d}\Phi (\bar t \rightarrow W^-\bar b){\rm d}\Phi
 (W^-\rightarrow \ell^-\bar \nu),   \label{shiki1}
\end{eqnarray}
where $s_x$ denotes the invariant mass of decaying particle $x$. 
In eq.(\ref{shiki1}), ${\rm d}\Phi$'s are as follows
\begin{eqnarray*}
 & & {\rm d}\Phi (e^+e^-\rightarrow t\bar t) = (2\pi)^4\frac{{\rm d}^3
{\bf p}_t}{(2\pi)^32E_t}\frac{{\rm d}^3{\bf p}_{\bar t}}{(2\pi)^3
2E_{\bar t}}\delta^{(4)}(P_{in}-p_t-p_{\bar t}),   \\
 & & {\rm d}\Phi (t\rightarrow W^+b) = (2\pi)^4\frac{{\rm d}^3
{\bf p}_{W^+}}{(2\pi)^32E_W}\frac{{\rm d}^3{\bf p}_b}
{(2\pi)^32E_b}\delta^{(4)}(p_t-p_{W^+}-p_b), 
\end{eqnarray*}
and so on, where $P_{in}, p_t, p_{\bar t}, p_{W^+}$ and $p_b$ are 
four-momenta of $e^+e^-$, $t, \bar t, W^+$ and $b$, respectively, 
${\bf p}_t, {\bf p}_{\bar t}, {\bf p}_{W^+}$ and ${\bf p}_b$ their 
three-momenta, and $E_t, E_{\bar t}, E_W$ and $E_b$ their energies. 
The relevant matrix element for a given helicity final state is 
\begin{eqnarray}
&M&\equiv (h_b, h_{\bar b}, h_{\ell^+}, h_{\ell^-}, h_{\nu}, 
h_{\bar \nu}) \equiv D_t(s_t)D_{\bar t}(s_{\bar t})D_{W^+}(s_{W^+})
D_{W^-}(s_{W^-})\sum_{h_t h_{\bar t}} (h_t,h_{\bar t}) \nonumber \\
& &\times\sum_{h_{W^+}}(h_t,h_{W^+},h_b)(h_{W^+},h_{\ell^+},  
h_{\nu})\sum_{h_{W^-}}(h_{\bar t},h_{W^-},h_{\bar b}) 
(h_{W^-},h_{\ell^-},h_{\bar \nu}),   \label{shiki2}
\end{eqnarray}
where $h_x$ denotes the helicity of particle $x$, $D_y$ the propagator of 
decaying particle $y$, and $(h_t,h_{\bar t})$, $(h_t,h_{W^+},h_b)$ and 
$(h_{w^+},h_{\ell^+},h_{\nu})$ are the helicity amplitudes for 
$e^+e^-\rightarrow t\bar t$, $t\rightarrow W^+b$ and $W^+\rightarrow 
\ell^+\nu$, respectively. In the above equation, the helicities of the initial 
$e^+e^-$ are omitted in the amplitude  $(h_t,h_{\bar t})$.

We apply to $D_y$ the narrow-width approximation for the real particle 
decays of $t, \bar t, W^+$ and $W^-$,
\begin{equation}
D_y(s_y) \simeq \frac{\pi }{m_y\Gamma_y}\delta(s_y-m_y^2),   
\label{shiki3}      
\end{equation}
where $m_y$ and $\Gamma_y$ are mass and total decay width of 
particle $y$. By integrating over ${\rm d}\Phi(\bar t \rightarrow 
W^-\bar b)$ and ${\rm d}\Phi(W^-\rightarrow \ell^-\bar \nu)$ 
in the differential cross section
\begin{equation}
   {\rm d}\sigma_{\rm tot} = |M|^2{\rm d}\Phi,            \label{shiki4}
\end{equation}
the differential cross section for $t\rightarrow W^+b\rightarrow 
\ell^+\nu b$ is given as 
\begin{eqnarray}
& &\frac{{\rm d}\sigma_{\rm tot}}{{\rm d}\Phi (t\rightarrow W^+b)
{\rm d}\Phi (W^+\rightarrow \ell^+\nu)} = \frac{{\rm Br}(\bar t 
\rightarrow W^-\bar b)}{2m_t\Gamma_t}\frac{{\rm Br}(W^-\rightarrow 
\ell^-\nu)}{2m_W\Gamma_W}{\rm d}\sigma (e^+e^-\rightarrow 
t\bar t)    \qquad  \nonumber  \\
& &\times \sum_{h_th'_t}\rho_{h_th'_t}\sum_{h_{W^+}}
(h_t,h_{W^+},h_b)(h_{W^+},h_{\ell^+},h_{\nu})\left[\sum_{h'_{W^+}}
(h'_t,h'_{W^+},h_b)(h'_{W^+}h_{\ell^+},h_{\nu})\right]^*,   \qquad
\label{shiki5}
\end{eqnarray}
where ${\rm Br}(\bar t\rightarrow W^-\bar b)$ and ${\rm Br}(W^-\rightarrow 
\ell^-\bar \nu)$ are the branching ratios for $\bar t \rightarrow 
W^-\bar b$ and $W^-\rightarrow \ell^-\nu$, respectively, 
${\rm d}\sigma (e^+e^-\rightarrow t\bar t)$ the differential cross 
section for $e^+e^-\rightarrow t\bar t$, and $\rho_{h_th'_t}$ is 
the normalized top-quark density matrix,
\begin{equation}
   \sum_{h_{\bar t}}(h_t,h_{\bar t})(h'_t,h_{\bar t})^* = \rho_{h_th'_t}
   \sum_{h_t,h_{\bar t}}|(h_t,h_{\bar t})|^2.             \label{shiki6}
\end{equation}

We introduce the up-down asymmetry of the lepton distributions from $W^+$ 
decays with respect to the $t\rightarrow W^+b$ decay plane in the top-quark 
laboratory frame for $e^+e^-\rightarrow t\bar t$, defined in \cite
{Grzad92} as 
\begin{equation}
         A^t \equiv \frac{N^t}{D^t} ,     \label{shiki7}  
\end{equation}
where 
\begin{eqnarray}
N^t&=&\int {\rm d}\Phi (t\rightarrow W^+b)\int_1^{-1}{\rm d}\cos\theta_
{\ell^+}\left[ \int_0^{\pi}-\int_{-\pi}^0\right] {\rm d}\phi_{\ell^+}
\frac{{\rm d}\sigma_{\rm tot}}{{\rm d}\Phi (t\rightarrow W^+b)
{\rm d}\Phi (W^+\rightarrow \ell^+\nu)},  \qquad  \label{shiki8} \\
D^t&=&\int {\rm d}\Phi (t\rightarrow W^+b)\int_1^{-1}{\rm d}\cos\theta_
{\ell^+}\left[ \int_0^{\pi}+\int_{-\pi}^0\right] {\rm d}\phi_{\ell^+}
\frac{{\rm d}\sigma_{\rm tot}}{{\rm d}\Phi (t\rightarrow W^+b)
{\rm d}\Phi (W^+\rightarrow \ell^+\nu)},   \qquad  \label{shiki9}
\end{eqnarray}
\vskip 0.3truecm
\noindent
where $\theta_{\ell^+}$ and $\phi_{\ell^+}$ are the polar and azimuthal 
angles of the lepton $\ell^+$ in the $W^+$ rest frame.

We adopt the most general parametrization of $tbW$ vertex for $t\rightarrow 
W^+b$ and $\bar t\rightarrow W^-\bar b$ decays as follows,
\begin{eqnarray}
\Gamma^{\mu}&=&-\frac{g}{\sqrt 2}V_{tb}\bar u (p_b)\left[ \gamma^
{\mu}(f_1^LP_L+f_1^RP_R)-\frac{{\rm i}\sigma^{\mu\nu}k_{\nu}}
{m_W}(f_2^LP_L+f_2^RP_R)\right] u(p_t),   \qquad  \label{shiki10}  \\
\bar \Gamma^{\mu}&=&-\frac{g}{\sqrt 2}V_{tb}^*\bar v (p_{\bar t})
\left[ \gamma^{\mu}(\bar f_1^LP_L+\bar f_1^RP_R)-\frac{{\rm i}\sigma^
{\mu\nu}k_{\nu}}{m_W}(\bar f_2^LP_L+\bar f_2^RP_R)\right] v(p_
{\bar b}),   \qquad  \label{shiki11} 
\end{eqnarray}
where $P_{L/R}=(1\mp \gamma_5)/2$, $k$ is the $W$ momentum, $V_{tb}$ 
is the $(tb)$-element of the Cabibbo-Kobayashi-Maskawa(CKM) mixing matrix 
and $g$ is the SU(2) gauge coupling constant. Since $W$ boson is on-shell, 
two other form factors do not contribute. The form factors of 
eqs.(\ref{shiki10}) and (\ref{shiki11}) are subject to [7,8]
\begin{equation}
    f_1^{L,R} = \pm \bar f_1^{L,R}, \qquad f_2^{L,R} = \pm \bar f_2^{R,L},   
\label{shiki12}
\end{equation}
where upper(lower) signs are those for CP-conserving(-violating) 
interactions.

If we use the amplitudes $(f_t,h_{W^+},h_b)$ for $t\rightarrow W^+b$ decay 
derived from the vertex functions (\ref{shiki10}) and (\ref{shiki11}) 
\cite{Grzad92}, the following expressions for $N^t$ and $D^t$ of eqs.(\ref
{shiki8}) and (\ref{shiki9}) arre obtained,  
\begin{eqnarray}
N^t&=&A2\pi^3\frac{m_t^2-m_W^2}{m_W^2}P^t_{\parallel}{\rm Im}
(f_1^Lf_2^{R*}),   \qquad  \label{shiki13}  \\
D^t&=&A\frac{16\pi^2}{3}\left[\frac{m_t^2+2m_W^2}{m_W^2}
|f_1^L|^2+6\frac{m_t}{m_W}{\rm Re}(f_1^Lf_2^{R*})+\frac{2m_t^2
+m_W^2}{m_W^2}|f_2^R|^2\right],   \qquad  \label{shiki14}
\end{eqnarray}
where $A$ is a common factor and $P^t_{\parallel}$ is the longitudinal 
polarization of the top-quark which occurs in the density matrix 
$\rho_{h_th_t'}$ of eq.(\ref{shiki6}) [6,20]. If we keep only the leading 
term in $D^t$, we obtain the following expression for the asymmetry 
$A^t$,
\begin{equation}
  A^t = h(m_t)P^t_{\parallel}{\rm Im}(f_1^Lf_2^{R*})/|f_1^L|^2,   
\label{shiki15}   
\end{equation}
where 
\[    h(m_t) = \frac{3\pi}{8}\frac{m_t^2-m_W^2}{m_t^2+2m_W^2}.   \]
\vskip 0.2truecm
\noindent
It is important to note that the CP-violation phase of the CKM matrix 
does not appear in the asymmetry $A^t$ and that $f_1^L = 1$ and 
$f_2^R = 0$ at the tree level and $f_2^R$ does not have any contribution 
even at the one-loop level in the Standard Model \cite{Grzad93}. 
So, $A^t$ is sensitive to the non-standard origin of CP violation.

Since $f_2^R$ has both CP-conserving and CP-violating parts, it is 
possible to have the only CP-violating contribution to the asymmetry by 
adding $A^{\bar t}$ for the anti-top decay $(\bar t\rightarrow W^-\bar b)$ to 
$A^t$ as follows,
\begin{equation}
      A \equiv A^t + A^{\bar t}.        \label{shiki16}      
\end{equation}
The relations of eq.(\ref{shiki12}) and the fact that $P^{\bar t}_{\parallel}
=-P^t_{\parallel}$ at the tree level for the $t$ and $\bar t$ production 
mechanism \cite{Kane92}\cite{Dalitz92} lead to 
\begin{equation}
   A = 2h(m_t)P^t_{\parallel}{\rm Im}(f_{2CPV}^{R*}),  \label{shiki17} 
\end{equation}
where $f_1^L=1$ at the tree level and $f_{2CPV}^R$ is the CP-violating 
contribution of $f_2^R$ given by 
\begin{equation}
   f_{2CPV}^R = \frac{1}{2}(f_2^R+\bar f_2^L).   \label{shiki18}   
\end{equation}
\vskip 1truecm

\leftline{\bf 3.The asymmetry $A^t$ in the two-Higgs doublet model}
\vskip 0.4truecm
\noindent
As stated in the previous section, since the asymmetry $A^t$ (or $A$) is 
sensitive to the non-standard origin of CP violation, we calculate this 
asymmetry in the simplest extension of the Standard Model, that is, 
in the two-Higgs doublet model.

The CP-violating neutral sector of two-Higgs doublet model is caused 
by the soft breaking in the Higgs potential of the discrete symmetry 
imposed on the Yukawa-coupling Lagragian. The discrete symmetry 
avoids the flavor-changing neutral current by coupling the charge 
$-\frac{1}{3}$ quarks and $+\frac{2}{3}$ quarks to the Higgs 
doublets $\phi_1$ and $\phi_2$, respectively (the so-called ^^ ^^ type II'' 
model), and their masses are generated through the vacuum expectation 
values $v_1$ and $v_2$ of $\phi_1$ and $\phi_2$, respectively. 
Relevant CP violation may be either explicit or spontaneous. 

We will not use a specific model but adopt the parametrization for the 
Higgs scalars defined by Weinberg\cite{Weinberg90}, which is formulated 
in unitary gauge through the unitarity gauge condition \cite{Weinberg73}, 
and Goldstone boson do not appear in this model. The three neutral scalars 
in the model are parametrized as 
\begin{equation}
\phi_1^0 = \frac{v_1}{\sqrt{2}|v_1|}\left[ \Phi_1-{\rm i}\frac
{|v_2|}{v}\Phi_3\right],  \qquad  \phi_2^0 = \frac{v_2}{\sqrt{2}|v_2|}
\left[ \Phi_2+{\rm i}\frac{|v_1}{v}\Phi_3\right]  \label{shiki19}
\end{equation}
where $v\equiv \sqrt{|v_1|^2+|v_2|^2}$, and the real new fields $\Phi_1, 
\Phi_2$ and $\Phi_3$ are subject to the canonical kinetic Lagrangian 
in unitary gauge,
\begin{equation}
   L_{\rm kin} = -\frac{1}{2}\sum_{n=1}^3(\partial_{\mu}\Phi_n)
(\partial^{\mu}\Phi_n).
\end{equation}
The two neutral scalars $\Phi_1$ and $\Phi_2$ are of CP-even and the 
third scalar $\Phi_3$ is of CP-odd, as is evident from eq.(\ref{shiki19}), 
so that CP violation shows up in the scalar exchange between quarks 
and/or gauge bosons through the imaginary patrs of the following four 
quantities\cite{Weinberg90},
\begin{equation}
  \frac{<\phi_2^0\phi_1^{0*}>}{v_1^*v_2},  \quad  \frac{<\phi_2^0
\phi_1^0>}{v_1v_2},  \quad  \frac{<\phi_1^0\phi_1^0>}{(v_1)^2},
\quad  \frac{<\phi_2^0\phi_2^0>}{(v_2)^2},       \label{shiki21}
\end{equation}
where $<\phi_i^0\phi_j^{0(*)}>$ denotes the propagator of neutral Higgs 
scalars (see Appendix). If we adopt the approximation of taking the 
effect of neutral scalar exchnage to be dominate by a single scalar 
particle of mass $m_{\phi}$, the four quantities are expressed as 
\cite{Weinberg90},\cite{Grzad92}
\begin{eqnarray}
  \frac{<\phi_2^0\phi_1^{0*}>}{v_1^*v_2} &=& \frac{\sqrt{2}G_FZ_0}
{m_\phi^2-q^2},  \qquad \frac{<\phi_2^0\phi_1^0>}{v_1v_2} = \frac{
\sqrt{2}G_F\tilde{Z}_0}{m_\phi^2-q^2},  \nonumber  \\
\frac{<\phi_1^0\phi_1^0>}{(v_1)^2} &=& \frac{\sqrt{2}G_FZ_1}
{m_\phi^2-q^2},  \qquad \frac{<\phi_2^0\phi_2^0>}{(v_2)^2} = 
\frac{\sqrt{2}G_FZ_2}{m_\phi^2-q^2},      \label{shiki22}
\end{eqnarray}
where we use the formalism of Feynman rules designed to minimize the 
number of times that the imaginary unit $i$ appears\cite{Aoki82}. 
By using eq.(\ref{shiki19}), Im$Z_0$, Im$\tilde{Z}_0$, Im$Z_1$ and 
Im$Z_2$ of eq.(\ref{shiki22}) are expressed as 
\begin{eqnarray}
  {\rm Im}Z_0&=&(1+\cot^2\beta)^{1/2}u_1u_3+(1+\tan\beta)^{1/2}
u_2u_3,   \nonumber   \\
  {\rm Im}\tilde{Z}_0&=&(1+\cot^2\beta)^{1/2}u_1u_3-(1+\tan^2\beta)
^{1/2}u_2u_3,   \nonumber   \\
  {\rm Im}Z_1&=&-2(\tan^2\beta+\tan^4\beta)^{1/2}u_1u_3,   \nonumber 
\\
  {\rm Im}Z_2&=&2(\cot^2\beta+\cot^4\beta)^{1/2}u_2u_3,   \label{shiki23}
\end{eqnarray}
where $\tan\beta=|v_2|/|v_1|$, and the three real numbers $u_1, u_2$ and 
$u_3$ are the coefficients of $\phi$ component of the $\Phi_1, \Phi_2$ 
and $\Phi_3$ states, respectively, and they are subject to the constraint, 
$u_1^2+u_2^2+u_3^2=1$. Only two of the four quantities of 
eq.(\ref{shiki23}) are independent and there are the following two relations 
among them,
\begin{eqnarray}
     & |v_1|^2{\rm Im}Z_1+|v_2|^2({\rm Im}\tilde{Z}_0+{\rm Im}Z_0)=0, 
\label{shiki24}  \\
     & |v_1|^2({\rm Im}\tilde{Z}_0-{\rm Im}Z_0)+|v_2|^2{\rm Im}Z_2=0. 
\label{shiki25}  
\end{eqnarray}

Now, we will compute the asymmetry $A^t$ in eq.(\ref{shiki15}). In this 
two-Higgs doublet model, we have $f_1^L=1$ at the tree level in 
the same way as in the Standard Model. The CP-violating part of the form 
factor $f_2^R$ can be obtained from the five one-loop diagrams of Fig.1, 
since we use the unitary gauge.

The relevant CP-violating propagators of the neutral scalars are the four 
quantities of eq.(\ref{shiki22}), $<\phi_2^0\phi_1^{0*}>, 
<\phi_2^0\phi_1^0>, <\phi_1^0\phi_1^0>, <\phi_2^0\phi_2^0>$ and 
their complex conjugates. The propagator of $W$ gauge boson in the 
unitary gauge is given as $(g_{\mu\nu}-\frac{q_{\mu}q_{\nu}}
{m_W^2-{\rm i}\varepsilon})/(q^2-m_W^2+{\rm i}\varepsilon)$ and 
that of the charged Higgs boson $H^+$ is given as $1/(m_H^2-q^2-
{\rm i}\varepsilon)$, where $m_H$ is the charged Higgs boson mass. 
The other neccesary Feynman rules for our calculation are summarized 
in the Appendix. The loop integrals for the diagrams in Fig.1 are 
expressed by the three-point functions defined in \cite{Passarino79}, 
\begin{eqnarray}
& &-\int \frac{{\rm d}^nq}{(2\pi)^n{\rm i}}\frac{\left[1, q_{\mu}, 
q_{\mu}q_{\nu}, q_{\mu}q_{\nu}q_{\alpha}\right]}{(q^2-m_1^2)\left[
(q+k)^2-m_2^2\right]\left[(q+k+p)^2-m_3^2\right]}  \nonumber   \\
& &=\frac{1}{(4\pi)^2}[ C_0, k_{\mu}C_{11}+p_{\mu}C_{12}, k_{\mu}
k_{\nu}C_{21}+p_{\mu}p_{\nu}C_{22}+(k_{\mu}p_{\nu}+p_{\mu}
k_{\nu})C_{23}+g_{\mu\nu}C_{24},   \nonumber  \\
& &k_{\mu}k_{\nu}k_{\alpha}C_{31}+p_{\mu}p_{\nu}p_{\alpha}C_{32}
+\{ pkk\}_{\mu\nu\alpha}C_{33}+\{ kpp\}_{\mu\nu\alpha}C_{34}
+\{ kg\}_{\mu\nu\alpha}C_{35}   \nonumber  \\
& &+\{ pg\}_{\mu\nu\alpha}C_{36} ], 
\label{shiki26}
\end{eqnarray}
where  $\{ pkk\}_{\mu\nu\alpha}=p_{\mu}k_{\nu}k_{\alpha}+k_{\mu}
p_{\nu}k_{\alpha}+k_{\mu}k_{\nu}p_{\alpha}, \{ kg\}_{\mu\nu\alpha}=
k_{\mu}g_{\nu\alpha}+k_{\nu}g_{\mu\alpha}+k_{\alpha}g_{\mu\nu}$.
All the $C$'s in eq.(\ref{shiki26}) have the arguments $C(k, p, m_1, 
m_2, m_3)$. In terms of these functions, we can calculate the 
CP-violating contributions to the form factor ${\rm Im}f_2^R$ for the 
five diagrams in Fig.1, consistently in the unitary gauge, as 
\begin{eqnarray}
{\rm Im}f_2^R|_1&=&\frac{1}{(4\pi)^2}\frac{gG_F}{v}m_b^2m_t[|v_1|^2
(-{\rm Im}Z_1+{\rm Im}\tilde{Z}_0)(C_{12}+C_{23})-|v_1|^2
{\rm Im}Z_0(C_0+C_{11})   \quad  \nonumber  \\
& &+|v_2|^2{\rm Im}Z_0(C_{11}+C_{21}-C_{12}-C_{23})],  \label{shiki27}
\end{eqnarray}
where $C\ldots = C\ldots (-p_t, p_{W^+}, m_b, m_H, m_{\phi})$,
\begin{eqnarray}
{\rm Im}f_2^R|_2&=&-\frac{1}{(4\pi)^2}\frac{gG_F}{v}m_t[ m_b^2|v_2|^2
(-{\rm Im}Z_2+{\rm Im}\tilde{Z}_0)(C_{12}+C_{23}-C_{11}-C_{21})
\quad  \nonumber  \\
& &+m_t^2|v_1|^2\{ ({\rm Im}\tilde{Z}_0-{\rm Im}Z_2)(C_0+C_{11})
+{\rm Im}Z_0(C_{12}+C_{23})\}],    \label{shiki28}
\end{eqnarray}
where $C\ldots = C\ldots (-p_b, -p_{W^+}, m_t, m_H, m_{\phi})$,
\begin{equation}
{\rm Im}f_2^R|_3 = -\frac{1}{(4\pi)^2}\frac{gG_F}{v}m_b^2m_t|v_2|^2
{\rm Im}Z_0(C_0+2C_{11}+C_{21}),   \qquad  \qquad   \label{shiki29}
\end{equation}
where $C\ldots = C\ldots (-p_t, p_{W^+}, m_b, m_W, m_{\phi})$,
\begin{eqnarray}
{\rm Im}f_2^R|_4&=&\frac{1}{(4\pi)^2}\frac{gG_F}{v}m_t|v_1|^2
{\rm Im}Z_0[ m_b^2C_{31}+(m_t^2-m_b^2-m_W^2)C_{33}+m_W^2C_{34}
-\frac{1}{6}   \quad  \nonumber  \\
& &+6C_{35}+2m_b^2C_{21}+m_W^2C_{22}+(m_t^2-2m_b^2-m_W^2)
C_{23}+4C_{24}  \nonumber  \\
& &+(m_b^2-2m_W^2)(C_{11}-C_{12})+m_t^2(C_{12}+C_{23})],  
\label{shiki30}
\end{eqnarray}
where $C\ldots = C\ldots (-p_b, -p_{W^+}, m_t, m_W, m_{\phi})$,
\begin{equation}
{\rm Im}f_2^R|_5=\frac{1}{(4\pi)^2}\frac{gG_F}{v}v^2m_b^2m_t[
{\rm Im}Z_0(C_{11}-C_{12})+{\rm Im}\tilde{Z}_0(C_{11}-C_{12}
+C_{21}-C_{23})],    \quad   \label{shiki31}
\end{equation}
where $C\ldots = C\ldots (-p_b, -p_{W^+}, m_{\phi}, m_b, m_t)$.

Now, we will evaluate the asymmetry $A^t$ in eq.(\ref{shiki15}). The free 
parameters are $u_1, u_2, u_3, \tan\beta (\equiv |v_2|/|v_1|), m_{\phi}$ 
and $m_H$. The neutral scalar boson mass $m_{\phi}$ has been restricted 
to $m_{\phi}=100-1000$ GeV \cite{APCTP96} or $m_{\phi}\leq 900$ GeV 
\cite{Dittmaier96} through the analyses of radiative corrections to $Z$ and 
$W$ boson masses in the Standard Model, and supersymmetric models 
constrain as $100 \leq m_{\phi} \leq 200$ GeV. ALEPH Collaboration 
has recently set a lower limit on the neutral Higgs boson mass of 63.9 GeV 
\cite{ALEPH96}. So, we choose $50 \leq m_{\phi} \leq 250$ GeV in our 
calculations, and we will take $100 \leq m_H \leq 1000$ GeV for the charged 
scalar boson mass. The dependence of the absolute value $|A^t|/
|P^t_{\parallel}|$ on the parameter $u_1, u_2$ and $u_3$ is obtained 
in Table 1 for $m_H=200$ GeV, $m_{\phi}=100$ GeV, $\tan\beta =1.0$ 
and top-quark mass of $m_t=180$ GeV. As seen from Table 1, the 
maximum value of the asymmetry $A^t$ proves to be of the order of 
$(1-2)\times 10^{-4}$, which is roughly consistent with the results of 
Grz\c{a}dkowski and Gunion \cite{Grzad92}.\footnote{They used the 
unitary gauge to determine the configuration of the neutral Higgs scalars. 
However, they mixed with it the 't Hooft-Feynman gauge for the 
propagators of charged scalars and $W$-boson and to draw the one-loop 
Feynman diagrams for $t\rightarrow W^+b$ decay.}
In the following calculations, we will take $u_1=-u_2=u_3=1/\sqrt{3}$, 
and assume $|P^t_{\parallel}|=1$ in order to estimate the maximum value 
of $A^t$. The magnitude of longitudinal polarization of $t$ quark 
depends on its production mechanism. The functions $C$'s develop 
imaginary parts for $m_H<m_t-m_b$ and always for the third diagram 
in Fig.1. In this case, ${\rm Im}f_2^R$ involves CP-conserving part of 
$Z$'s, that is, ${\rm Re}Z_0, {\rm Re}\tilde{Z}_0$, etc., and we do not 
include this part for the calculation of CP-violating ${\rm Im}f_2^R$. 

\begin{table}
\begin{center}
\begin{tabular}{|rrr|c|c|}    \hline
$\ (u_1,$ & $u_2,$ & $u_3)\ $ & $\ {\rm Im}f_2^R \ [\times 10^{-3}]\ $ &
$\ |A^t|/|P^t_{\parallel}|\ [\times 10^{-3}]\ $ \\   \hline
$1/\sqrt{2}$ & 0 & $1/\sqrt{2}$ & $-0.16$ & $0.11$   \\
$1/\sqrt{2}$ & 0 & $-1/\sqrt{2}$ & $0.16$ & $0.11$   \\
$0$ & $1/\sqrt{2}$ & $1/\sqrt{2}$ & $0.35$ & $0.24$  \\
$0$ & $1/\sqrt{2}$ & $-1/\sqrt{2}$ & $-0.35$ & $0.24$  \\
$1/\sqrt{3}$ & $1/\sqrt{3}$ & $1/\sqrt{3}$ & $0.13$ & $0.084$  \\
$1/\sqrt{3}$ & $1/\sqrt{3}$ & $-1/\sqrt{3}$ & $-0.13$ & $0.084$  \\
$1/\sqrt{3}$ & $-1/\sqrt{3}$ & $1/\sqrt{3}$ & $-0.34$ & $0.23$  \\
$1/\sqrt{3}$ & $-1/\sqrt{3}$ & $-1/\sqrt{3}$ & $0.34$ & $0.23$  \\  \hline
\end{tabular}
\end{center}
\caption{Dependence of ${\rm Im}f_2^R$ and $|A^t|/|P^t_{\parallel}|$ 
on the parameter set $(u_1, u_2, u_3)$ under the constraint of 
$u_1^2+u_2^2+u_3^2=1$.}
\label{tab1}
\vskip 0.1truecm
\end{table}

The numerical results are shown in Figs.2 and 3. As seen in Fig.2, $m_t$-
dependence of $|A^t|$ is strong, as expected from its dependence of Higgs 
scalar coupling to the fermions. On the contrary, as seen in Fig.3, 
the dependence on the neutral and charged scalars masses $m_{\phi}$ and 
$m_H$ is moderate, since their dependence is only of ${\rm log}(m_{\phi})$ 
and ${\rm log}(m_H)$ in the one-loop amplitudes. Fig.4 shows the 
dependence of $|A^t|$ on $\tan\beta$. Consequently, the CP-violating 
up-down asymmetry of $W$-decaying leptons in $t\rightarrow W^+b$ decay 
is of the order of $10^{-3}-10^{-5}$ for $0.5\leq \tan\beta \leq 10$ for 
reasonable Higgs scalars mass range, $50 \leq m_{\phi}\leq 250$ GeV and 
$100 \leq m_H \leq 1000$ GeV.
\vskip 1truecm

\leftline{\bf 4. Conclusions and discussions}
\vskip 0.4truecm
\noindent
We have investigated CP violation in the top-quark decay by studying the 
asymmetry ($A^t$) of lepton distributions from the subsequent decay 
$W^+\rightarrow \ell^+\nu$ in $t\rightarrow W^+b$ in the two-Higgs 
doublet model with CP-violating neutral sector. As expected, due to the 
large coupling of Higgs scalars to the top-quark, the asymmetry is 
significantly large $((1-3)\times 10^{-4})$ for the one-loop effects, 
for the typical parameter values of $m_{H^+}=200$ GeV, $m_\phi=100$ 
GeV and $\tan\beta=1.0$, though this magnitude of the asymmetry 
may be hard to detect by the experiments. This value goes up to 
$1\times 10^{-3}$ for $\tan\beta =0.5$ for the same $m_H$ and $m_\phi$ 
values. 

If CP-conserving contributions are included, this asymmetry shows values 
$(\sim 10^{-3})$ larger by one order of magnitude, as seen in Fig.5. It is 
due to the third diagram of Fig.1 which develops a large imaginary part 
for the vertex amplitude. Therefore, the magnitude of CP-violating 
contribution to the asymmetry $A^t$ is about $1/10$ of the CP-conserving 
contribution. Experimentally, the CP-violating contribution could be 
extracted by adding the asymmetries from $t$ and $\bar t$ decays 
as stated in the section 2. 
\vskip 1truecm

\centerline{ACKNOWLEDGEMENTS}
\vskip 0.4truecm

We are grateful to Z. Hioki for useful discussions.
\vskip 1.5truecm
\newpage
\noindent
{\bf Appendix}
\vskip 0.4truecm
\noindent
In this Appendix we summarize the Feynman rules which are neccesary 
for the calculations in this paper in the two-Higgs doublet model with 
CP-violating neutral sector. The model we adopted here is the so-called 
"type II", in which one Higgs doublet $\phi_1^0$ couples exclusively to 
the "down"-type quarks and another doublet $\phi_2^0$ to the "up"-type 
quarks. Our formalism for constructing the Feynman rules is the one 
designed to minimize the number of times that the imaginary unit $i$ 
appears\cite{Aoki82}. In the following, we denote the $tbW$-vertex, 
for example, as $t\bar b W^-_\mu$, which means that all of the three 
particles $t, \bar b$ and $W^-$ enter the vertex, and the Feynman rule 
for this vertex is $g/\sqrt{2}\gamma_\mu P_L$ according to our formalism. 

Quark couplings to neutral Higgs scalars;
\begin{eqnarray*}
D_j\bar D_j\phi_1^0 &:& -\frac{1}{v_1}m_{D_j}P_R,  \qquad  D_j\bar D_j
\phi_1^{0*} : -\frac{1}{v_1^*}m_{D_j}P_L,  \\
U_i\bar U_i\phi_2^0 &:& -\frac{1}{v_2}m_{U_i}P_L,  \qquad  U_j\bar U_i
\phi_2^{0*} : -\frac{1}{v_2^*}m_{U_i}P_R,  
\end{eqnarray*}
where $D_j$ and $U_i$ are "down"-type quark and "up"-type quark, 
respectively, $m_{D_j}$ and $m_{U_i}$ their masses, $P_{R/L}=(1\pm 
\gamma_5)/2$, and $v_1$ and $v_2$ are vacuum expectation values for 
$\phi_1^0$ and $\phi_2^0$, respectively.

Quark couplings to charged Higgs scalar $H^{\pm}$;
\begin{eqnarray*}
U_i\bar D_jH^- &:& \frac{1}{v}V_{ij}^*(\frac{v_2}{v_1^*}m_{D_j}P_L
+\frac{v_1}{v_2^*}m_{U_i}P_R),  \\
D_j\bar U_iH^+ &:& \frac{1}{v}V_{ij}(\frac{v_2^*}{v_1}m_{D_j}P_R
+\frac{v_1^*}{v_2}m_{U_i}P_L), 
\end{eqnarray*}
where $v\equiv \sqrt{|v_1|^2+|v_2|^2}$ and $\sqrt{2}G_F=1/(2v^2)$.
$V_{ij}$ is the $(ij)$-element of Cabibbo-Kobayashi-Maskawa mixing matrix.
$H^{\pm}$ are the charged Higgs scalars. Goldstone bosons do not appear 
in our formalism, since we use the unitary gauge.

Gauge boson pair couplings to neutral Higgs scalars;
\[  W^+_{\mu}W^-_\nu\phi_{1,2}^0 : \frac{1}{2}g^2v_{1,2}^*g_{\mu\nu},
 \qquad W^+_{\mu}W^-_\nu\phi_{1,2}^{0*} : \frac{1}{2}g^2v_{1,2}g_{\mu\nu},
\]
where $g$ is the SU(2) gauge coupling constant and $g=\sqrt{2}M_W/v$. 

Gauge boson couplings to charged Higgs scalar and neutral Higgs scalar;
\begin{eqnarray*}
W_{\mu}^+H^-\phi_1^0 &:& \frac{g}{\sqrt{2}}\frac{v_2}{v}(p-q)_\mu, \qquad
W_{\mu}^-H^+\phi_1^{0*} : -\frac{g}{\sqrt{2}}\frac{v_2^*}{v}(p-q)_\mu, \\
W_{\mu}^+H^-\phi_2^0 &:& -\frac{g}{\sqrt{2}}\frac{v_1}{v}(p-q)_\mu, \qquad
W_{\mu}^-H^+\phi_2^{0*} : \frac{g}{\sqrt{2}}\frac{v_1^*}{v}(p-q)_\mu, 
\end{eqnarray*}
where $p$ and $q$ are the momenta of the incoming $H^{\pm}$ and 
$\phi_{1,2}^{0(*)}$, respectively.

We should mention that the propagator $<\phi_i^0\phi_j^0>$ means that 
$\phi_i^0$ and $\phi_j^0$ are both entering the vertices and directed 
back to back on the propagator.

\newpage
\centerline{\bf Figure captions}

\vskip 0.5truecm
\noindent
{\bf Fig.1.} The five one-loop diagrams contributing to the CP-violating 
part of the form factor $f_2^R$ in $t\rightarrow W^+b$ decay in the 
two-Higgs doublet model in the unitary gauge. The propagators with 
$\phi^0$ are the ones of $<\phi_i^0\phi_j^{0(*)}>$ of CP-violating 
neutral Higgs-boson exchange.

\vskip 0.3truecm
\noindent
{\bf Fig.2.} The dependence of the maximum absolute value of the 
asymmetry $A^t$ on the top-quark mass $m_t$ for various sets of charged 
and neutral scalar boson masses; $(m_H,m_{\phi}=(100,50)$GeV 
(dashed-dotted), $(200,100)$GeV (solid), $(500,100)$GeV (dashed) 
and $(1000,100)$GeV (dotted curve).

\vskip 0.3truecm
\noindent
{\bf Fig.3.} The dependence of the maximum absolute value of the 
asymmetry $A^t$ on the neutral scalar boson mass $m_\phi$ for the charged 
scalar boson mass, $m_H=100$GeV (solid), 200GeV (dashed-dotted), 
500GeV (dashed) and 1000GeV (dotted curve).

\skip 0.3truecm
\noindent
{\bf Fig.4.} The dependence of the maximum absolute value of the 
asymmetry $A^t$ on the ratio of vacuum expectation values of the two 
neutral Higgs fields, $\tan\beta (\equiv |v_2|/|v_1|)$, for $(m_H,m_\phi)=
(200,100)$GeV.

\vskip 0.3truecm
\noindent
{\bf Fig.5.} The dependence of the maximum absolute value of the 
asymmetry $A^t$ with the inclusion of CP-conserving contributions to the 
form factor $f_2^R$ on the neutral scalar boson mass $m_\phi$ for the 
charged scalar boson mass, $m_H=100$GeV (solid), 200GeV (dashed-dotted), 
500GeV (dashed) and 1000GeV (dotted curve).

\end{document}